\documentclass[journal=jpcbfk, manuscript=article, layout=twocolumn]{achemso}
\setkeys{acs}{email=false}

\usepackage{amsmath, amssymb}
\usepackage{graphicx}
\usepackage{booktabs}
\usepackage{capt-of}
\usepackage{hyperref}
\hypersetup{hidelinks}
\usepackage{xcolor}
\usepackage{cuted}
\usepackage{placeins}
\usepackage{dblfloatfix}

\usepackage{multirow}
\usepackage{array}


\title{A Machine Learning Framework for Magnetic Candidate
       Discovery in Cerium-Based Compounds}

\author{Joshua A. Torres$^{*,\ddagger}$, Yaser M. Banad$^{\ddagger}$,
        Benjamin O. Tayo$^{\dagger}$, and Tej Nath Lamichhane$^{*,\dagger}$}
\affiliation{$^{\dagger}$School of Engineering,
             University of Central Oklahoma, Edmond, Oklahoma 73034,
             United States\\
             $^{\ddagger}$School of Electrical and Computer Engineering,
             University of Oklahoma, Norman, Oklahoma 73019,
             United States\\[0.6em]
             E-mail: jatorres@ou.edu; tlamichhane@uco.edu}

\abbreviations{ML, DFT, GK, FM, AFM, RF, SOAP, MC}


\begin{document}
% =============================================================

% ---- abstract -----------------------------------------------
\begin{strip}
\begin{minipage}{\textwidth}
\noindent{\Large\bfseries Abstract}\par\vspace{0.6em}
\small
Cerium (Ce), the most abundant lanthanide, offers significant potential
for developing sustainable magnetic materials.
Predicting ferromagnetic compounds with strong uniaxial magnetic
anisotropy remains challenging because magnetic ordering depends
sensitively on crystal structure, exchange geometry, and electronic
interactions.
Here, we present a physics-guided computational framework that
integrates Random Forest (RF)-based structural screening, a Ce-specific
modification of the Goodenough--Kanamori exchange model, Ising-model
Monte Carlo simulations, and unsupervised representation learning to
screen known Ce-based crystal structures, identify promising candidates
exhibiting Ising-like ferromagnetic behavior, and characterize their
magnetic phase transitions and critical properties.

The top candidate structures are simulated on their crystal-derived
magnetic lattices to map phase behavior across temperature and
magnetization.
Phase diagrams obtained from these simulations are used to extract
critical exponents, enabling quantitative characterization of magnetic
regimes and phase transitions.
To further strengthen the predictive framework, we analyze the simulated
configurations using an autoencoder-based unsupervised learning
approach.
The resulting latent representations capture signatures of phase
evolution and transition behavior, providing an interpretable connection
between simulated spin configurations and emergent magnetic phenomena.

Together, these methods establish a scalable framework that combines
structural screening, statistical-mechanical simulation, critical
exponent extraction, and unsupervised learning in the search for new
magnetic materials.
Beyond classification alone, this approach advances the physics-informed
filtering of known crystals and provides a pathway for identifying
candidate compounds for future synthesis and experimental validation.
Ultimately, this work deepens understanding of magnetism in
cerium-based compounds and offers a broader platform for exploring
next-generation magnetic materials across the lanthanide and actinide
families.
\end{minipage}
\end{strip}

% Graphical abstract: fixed at the top of the next page before Introduction text.
\twocolumn[{%
  \begin{minipage}{\textwidth}
    \centering
    \noindent{\Large\bfseries Graphical Abstract}\par\vspace{0.5em}
    \includegraphics[width=\textwidth]{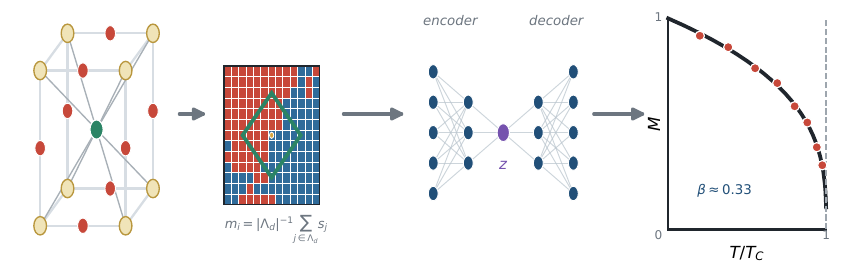}
    \par\vspace{1em}
  \end{minipage}
}]

\begin{figure*}[tp]
  \centering
  \includegraphics[width=0.92\textwidth]{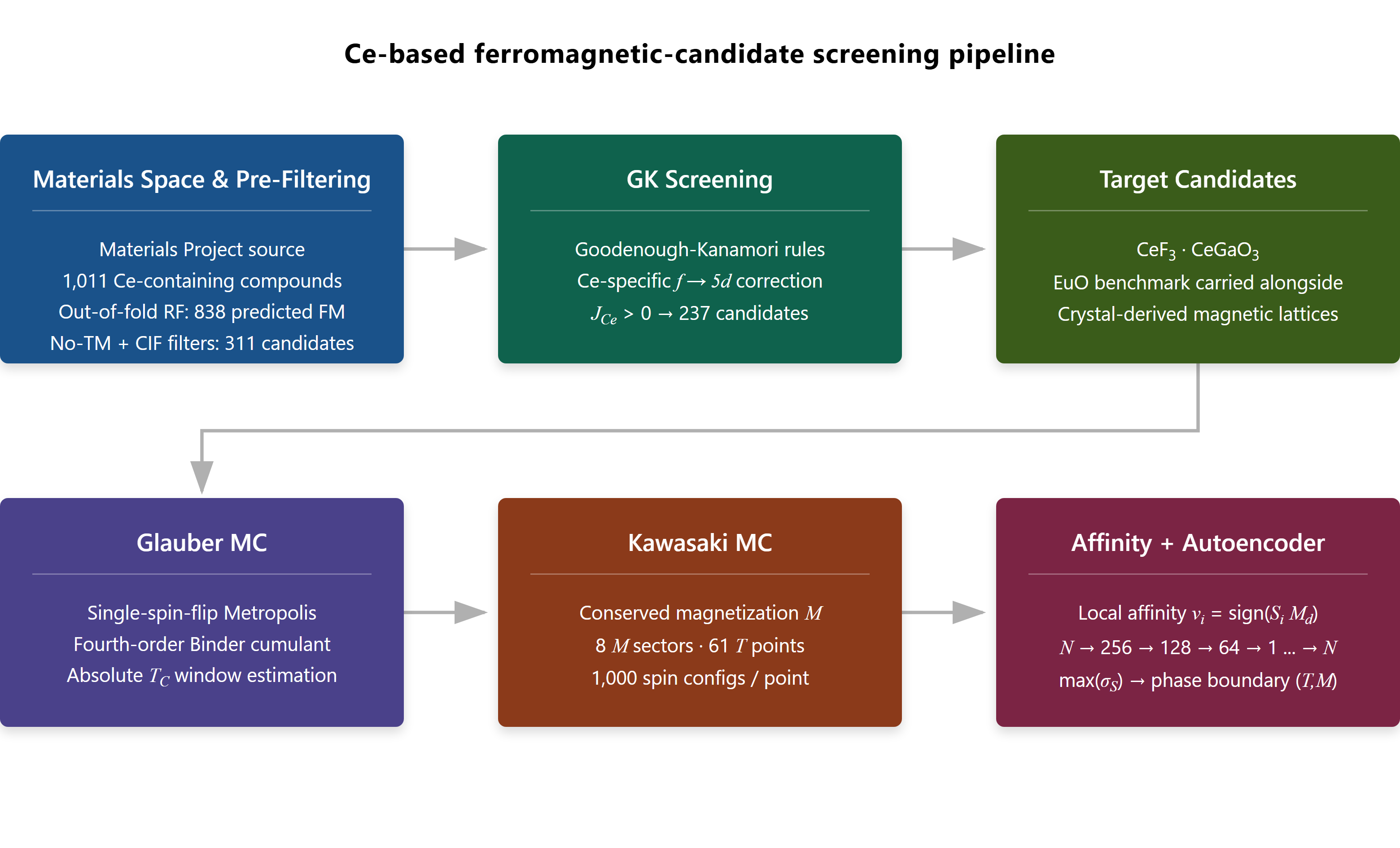}
  \caption{Ce-based ferromagnetic-candidate screening and simulation
    workflow. A Random Forest trained on Materials Project magnetic-order
    labels produces out-of-fold structural predictions, which are filtered
    by modified Goodenough--Kanamori screening. The retained candidates are passed through Glauber and
    Kawasaki Monte Carlo simulations, and local-affinity vectors are
    analyzed by an autoencoder to reconstruct phase boundaries.}
  \label{fig:workflow}
\end{figure*}

% =============================================================
\section{Introduction}
% =============================================================

Ferromagnetic materials occupy a privileged role in quantum
materials physics because long-range spin order enables magnetic
functionality in spintronic, magnetocaloric, and quantum-materials
platforms.\cite{Zutic2004,Pecharsky1999,Coey2010}
Despite their technological promise, systematic computational
discovery of new cerium-based ferromagnetic candidates remains limited.
The target materials in this work are not merely ferromagnets, but
ferromagnets with sufficiently strong uniaxial magnetic anisotropy to
motivate an effective Ising-like description.  Accordingly, the
Monte Carlo stage below should be understood as a coarse-grained
phase-mapping model for candidate magnetic graphs; direct confirmation
of easy-axis anisotropy remains a follow-up electronic-structure or
experimental validation step.

Cerium presents an exceptional opportunity in this space.
The Ce$^{3+}$ ion carries a $4f^1$ configuration, so magnetic coupling
in Ce compounds need not be restricted to a single microscopic
mechanism; the dominant exchange can vary with bonding, coordination,
and electronic structure.  In this work we focus specifically on the
anion-mediated Ce--anion--Ce pathways that can be extracted
consistently from CIF structures and screened using
Goodenough--Kanamori (GK) rules.\cite{Goodenough1963,Kanamori1959}
Studies of rare-earth monochalcogenides further show that exchange
between localized $4f$ moments can be affected by virtual excitations
into unoccupied rare-earth $5d$ levels,\cite{Kasuya1968,Kunes2005}
while Ce compounds additionally exhibit significant $4f$--ligand
hybridization and crystal-field effects that modify the exchange
interaction.\cite{Jo1993}
This motivates the Ce-specific screening correction used here: an
analogous $f \!\to\! 5d$ virtual channel is added to the GK estimate for
Ce--anion--Ce pathways (hereafter referred to as modified-GK) before
candidates are passed to Monte Carlo validation.
Beyond its abundance and low cost, Ce exhibits rich magnetic behavior
arising from competing $4f$--$5d$ interactions and strong sensitivity to
crystal structure and chemical bonding.
Compared with the extensively investigated Nd- and Sm-based magnet
systems, many Ce compounds remain comparatively unexplored, providing
significant opportunities for computationally guided materials
discovery.

Machine learning has recently emerged as a transformative tool in
computational materials discovery.\cite{Bedolla2021}
Supervised classifiers trained on structural and electronic descriptors
have been used as fast surrogates for magnetic-order prediction in
correlated $f$-electron materials,\cite{Ghosh2020,Broyles2024} while
unsupervised approaches --- notably
autoencoders --- have been used to map order parameters and phase
boundaries directly from simulation snapshots without prior
knowledge of the transition.\cite{Wang2016,Jang2025}
The combination of supervised structural benchmarking, physics-based
candidate screening, and unsupervised phase-diagram construction forms
the conceptual backbone of the present work.
Accordingly, the manuscript follows this sequence: define the
Ce-containing materials space, benchmark structural ferromagnetic
prediction, apply the modified-GK positive-exchange screen, estimate
temperature windows with Glauber MC, reconstruct Kawasaki phase
boundaries with local affinities and an autoencoder, and finally extract
critical exponents from the completed finite-size data.
Here we adapt the constant-magnetization Ising protocol of Jang and
Yethiraj\cite{Jang2025} to material-derived Ce magnetic graphs obtained
from CIF structures and modified-GK screening.

Several ML studies of magnetic ordering exist, but most address
$3d$ transition metal oxides or focus exclusively on the critical
point of the Ising model.\cite{Wang2016,Jang2025}
We are not aware of any systematic ML-based screening of $4f$
compounds that propagates candidates through a full Monte Carlo
phase-diagram pipeline.
The goal of this work is precisely to close that gap: we build and
validate a workflow that combines an RF structural-screening stage
with GK screening, Glauber and Kawasaki MC simulation,
and autoencoder phase-diagram analysis specifically tuned for
Ce-based ferromagnetic candidates that could later be prioritized for
uniaxial-anisotropy validation, and we
demonstrate it against EuO as an external rare-earth ferromagnetic
benchmark for the MC/autoencoder workflow.\cite{Kasuya1968,Mauger1986}

% =============================================================
\section{Model and Methods}
% =============================================================

\subsection{Materials Space and Data Collection}

We query the Materials Project API\cite{Jain2013,Ong2015} for all
Ce-containing binary and ternary crystal structures. Crystal structure
files (CIF format) are downloaded and converted to pymatgen
\texttt{Structure} objects\cite{Ong2013} for subsequent analysis.
Figure~\ref{fig:phase_mapping_workflow} summarizes the CIF-to-autoencoder workflow.

\subsection{Ferromagnetic Prediction Using Structural Characteristics}

\begin{figure*}[tp]
  \centering
  \includegraphics[width=0.98\textwidth]{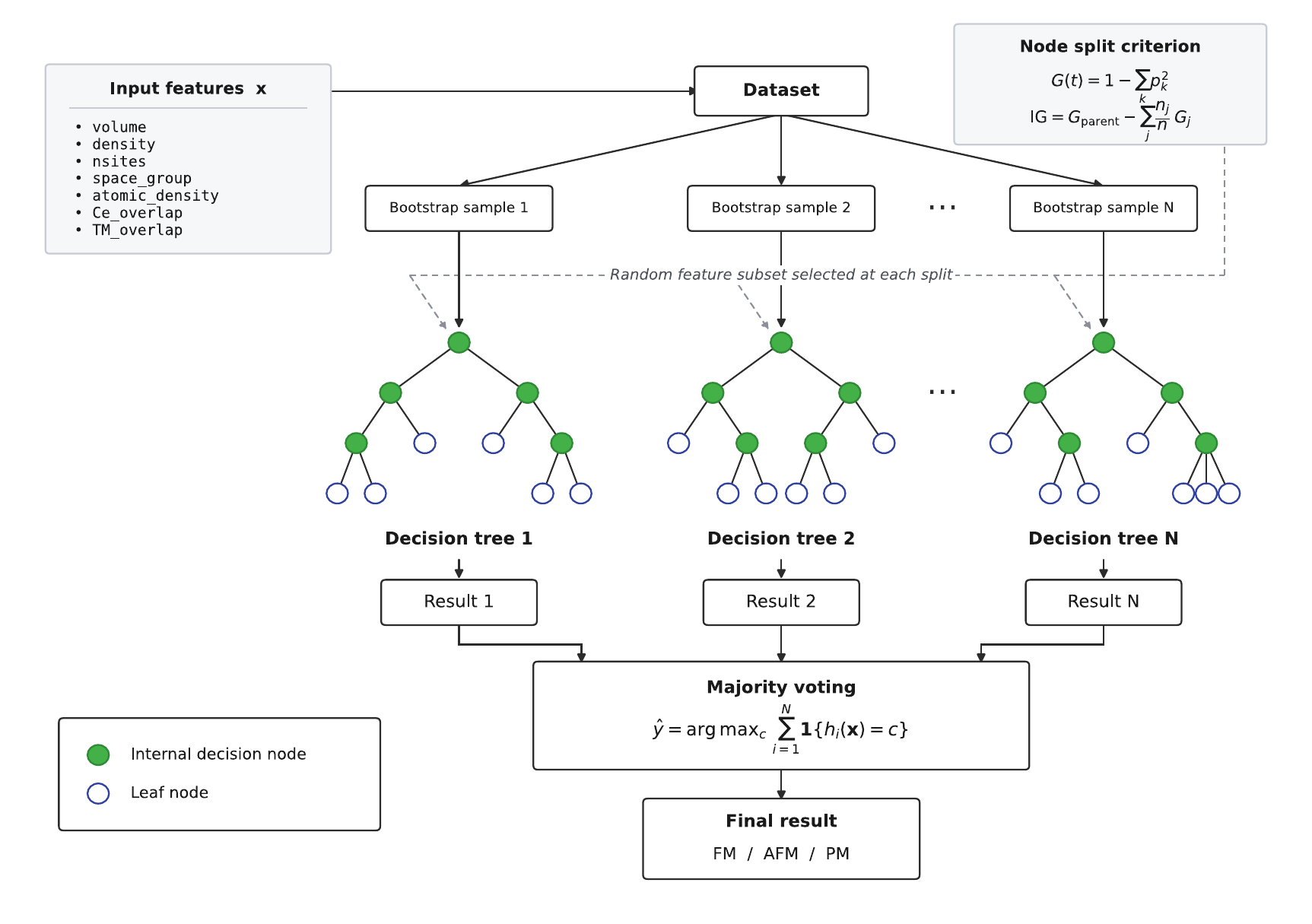}
  \caption{Random Forest classifier schematic used for
    ferromagnetic-label prediction from structural and SOAP-derived
    descriptors. Individual decision trees produce independent class
    votes, and the ensemble prediction is assigned by majority voting.}
  \label{fig:random_forest}
\end{figure*}

Following Broyles and colleagues,\cite{Broyles2024} we adopt the same
three-class RF formulation (FM, AFM, and PM), but apply it here to
Ce-containing compounds labeled from Materials Project magnetic
ground-state annotations rather than the experimentally curated uranium
dataset used in the original study.
As in the reference work, each compound is represented by a compact
descriptor constructed from purely structural information, allowing
magnetic-order prediction without explicit DFT-derived inputs.
In the present implementation, this descriptor combines bulk structural
features with Smooth Overlap of Atomic Positions (SOAP)
overlaps.\cite{Bartok2013}
Figure~\ref{fig:random_forest} summarizes the RF ensemble.

The compact descriptor contains seven features: the unit-cell volume,
mass density, number of atomic sites, space group, atomic density
(sites per unit volume), a Ce SOAP overlap, and a transition-metal
(TM) SOAP overlap.
The SOAP power spectra are computed with the DScribe package using
$n_{\max} = 5$, $l_{\max} = 3$, and a cutoff radius
$r_c = 5$\,\AA{}, and are then reduced to per-species scalar overlaps by
summing over the angular and radial channels and over atomic sites.
This construction preserves a low-dimensional representation of the
crystal structure while retaining information about the local
Ce-centered coordination environment and the degree of TM participation.

Three dataset scopes are evaluated: a binary set of Ce compounds with
exactly two distinct elements ($N = 208$), a ternary set
($N = 803$), and the combined set ($N = 1{,}011$).
All three datasets are class-imbalanced and are dominated by the FM
label.
In the binary set, the class distribution is 165 FM, 18 AFM, and
25 PM compounds, corresponding to $79.3\%$, $8.7\%$, and $12.0\%$ of
the dataset, respectively.
In the ternary set, the distribution is 556 FM, 130 AFM, and
117 PM compounds ($69.2\%$, $16.2\%$, and $14.6\%$), while the combined
set contains 721 FM, 148 AFM, and 142 PM compounds
($71.3\%$, $14.6\%$, and $14.0\%$).
The strong FM skew is therefore a central feature of all three
classification problems and must be considered when interpreting model
performance.

To benchmark the descriptor under the same general RF framework used by
Broyles and colleagues,\cite{Broyles2024} we train a Random Forest with
100 decision trees, Gini-impurity splitting, and a maximum of 10 features
considered at each split. For the compact seven-feature descriptor this
bound exceeds the descriptor dimension, so all seven features are
available at every split and tree-to-tree variation arises from bootstrap
sampling alone; for the abundance-augmented representation it acts as a
genuine random feature subset, as in the reference implementation.
Performance is estimated over 10{,}000 independent stratified
$70/30$ train--test splits for each dataset scope.
For the seven-feature structure-plus-SOAP descriptor, the stochastic-split
mean accuracies are $\mu = 79.3 \pm 3.7\%$ for the binary set,
$\mu = 69.0 \pm 2.1\%$ for the ternary set, and
$\mu = 71.0 \pm 1.8\%$ for the combined set.
On representative fixed splits, the corresponding balanced accuracies
are 0.590, 0.434, and 0.453, with macro-$F_1$ scores of
0.615, 0.450, and 0.467 for the binary, ternary, and combined sets,
respectively.
These results show that overall accuracy alone overstates performance
because it is influenced by the majority FM class.
The more conservative balanced-accuracy and macro-$F_1$ metrics indicate
that the descriptor captures the dominant FM signal more reliably than
the minority AFM and PM classes, which remain more difficult to resolve
under the present class imbalance.
The binary problem gives the strongest baseline performance, whereas the
ternary and combined sets are more challenging because the label space
must be learned across a broader and more compositionally diverse
materials set.

We also tested an expanded representation that augments the seven
structural features with elemental-abundance inputs.
This abundance-augmented model yields the highest overall performance,
with stochastic-split mean accuracies of $\mu \approx 82.0\%$ for the
binary set, $\mu \approx 74.2\%$ for the ternary set, and
$\mu \approx 76.1\%$ for the combined set.
The improvement indicates that compositional weighting contributes
useful predictive information beyond the compact structural-plus-SOAP
descriptor alone, especially for the more data-rich ternary and combined
datasets.

Taken together, these RF results provide a rapid structural benchmark for
magnetic-label prediction in Ce compounds and show that
the feature space adapted from Broyles and colleagues remains
informative when transferred from uranium-based systems to Ce-based
chemistries.
To create the candidate pool without predicting a training structure with
a model that has already seen its label, we apply repeated stratified
10-fold cross-validation to the combined dataset.  Each compound is held
out and predicted in ten independent folds, and its mean class
probabilities determine the out-of-fold RF assignment.  This procedure
predicts 838 of the 1,011 compounds as FM.  These RF predictions, rather
than the Materials Project training labels themselves, provide the input
to the modified-GK screening funnel described below.

\subsection{Goodenough--Kanamori Screening with $f$-Orbital Correction}

Standard GK rules classify superexchange interactions based on the
occupancy and geometry of overlapping $d$ orbitals on adjacent magnetic
sites.\cite{Goodenough1963,Kanamori1959}
We use this framework as a high-throughput exchange screen rather than
as a fully \textit{ab initio} evaluation of the superexchange integral:
the GK rules identify the active orbital/geometric channels and their
FM or AFM signs, while Kanamori-style transfer-integral expressions
provide semi-empirical relative magnitudes.  This approximation is
used as a screening-level model for rare-earth compounds because
exchange between localized $4f$ moments can acquire additional
contributions from virtual excitations into unoccupied rare-earth
$5d$ levels, as established in Eu chalcogenide literature and
first-principles Eu monochalcogenide studies.\cite{Kasuya1968,Kunes2005}
The Eu literature motivates the physical form of the screening
correction but is not used to parameterize the Ce-specific model.
For Ce, the relevant occupied orbital is $4f$ rather than $3d$, and
the exchange estimate is further modified by low-lying $5d$ virtual
states accessible through crystal-field mixing.  We implement this as
a Ce-specific additive channel model for each Ce--anion--Ce pathway,
\begin{equation}
    J_p = \sum_c J_{p,c}
        = J_{\sigma} + J_{\pi} + J_{90^\circ} + J_{f\to5d},
    \label{eq:GK}
\end{equation}
where the standard $\sigma$, $\pi$, and $90^\circ$ terms follow the
GK sign rules and the additional $f\!\to\!5d$ term is a positive
virtual-excitation channel of the form
$J_{f\to5d}\propto 2b_{fd}^2/[(4S_iS_j)\Delta_{fd}]$, scaled by the
pathway angular factor.  The material-level screening descriptor is
the mean over nonzero pathway sums, $J_{\mathrm{Ce}}=\langle
J_p\rangle_{J_p\ne0}$.  Materials with $J_{\mathrm{Ce}}>0$ are retained
as FM candidates; the remainder are classified as AFM or nonmagnetic
within the screening model.
The modified-GK values are used both to identify positive-exchange Ce
graphs and, through their material-level mean, as the uniform coupling
for the Glauber and Kawasaki models. Pathway-to-pathway variations are
not inserted into the present Hamiltonian.

The modified-GK screen is applied after the RF stage has identified the
predicted ferromagnetic candidate pool. Starting from 1{,}011
Ce-containing compounds, the out-of-fold RF screen predicts 838 as FM.
Requiring zero transition-metal SOAP overlap narrows this set to 316
compounds dominated by pure Ce--anion--Ce pathways, and CIF availability
leaves 311 structures for explicit modified-GK analysis. Applying the
Ce-specific $f \!\to\! 5d$ channel model and retaining materials with
$J_{\mathrm{Ce}}>0$ yields 237 positive-exchange screening candidates
for downstream prioritization.

\subsection{Monte Carlo Simulation with Conserved Magnetization}

For each screened candidate we construct a supercell with the
magnetic lattice extracted from the crystal structure.
Following the constant-magnetization Ising protocol of Jang and
Yethiraj,\cite{Jang2025} the Ising model has $N$ two-state spins on a
lattice, with one coarse-grained orientation variable assigned to each
magnetic site.
The Hamiltonian is
\begin{equation}
    H = -J \sum_{\langle ij \rangle} S_i S_j,
    \label{eq:ising}
\end{equation}
where $S_i \in \{+1,-1\}$ is the spin on site $i$ and
$\langle ij\rangle$ denotes nearest-neighbor magnetic pairs.
In the present materials workflow, these nearest-neighbor pairs are
identified from the CIF-derived magnetic graph.
Here $S_i$ encodes the orientation of the local moment and should not
be interpreted as the microscopic spin quantum number of the ion.
The modified-GK value is used for screening, for identifying the
magnetic Ce--anion--Ce connectivity, and as the uniform material-level
coupling $J=J_{\mathrm{GK}}$ on every retained ferromagnetic bond. The
Monte Carlo calculation therefore tests the phase behavior of a
uniformly weighted exchange/connectivity graph using the mean
modified-GK energy scale. Magnetocrystalline anisotropy remains a
separate materials-validation criterion, and pathway-resolved GK
magnitudes are not inserted into the present Kawasaki Hamiltonian.
All temperatures are reported in kelvin, with the material-level coupling
expressed in meV and $k_B$ in meV K$^{-1}$. To fix the absolute
temperature range for each candidate, we first
estimate its critical temperature $T_C$ using non-conserved Glauber
single-spin-flip dynamics,\cite{Glauber1963} locating $T_C$ from the crossing of the
fourth-order Binder cumulant across system sizes. This provides
$T_C$ in kelvin and sets the window over which the conserved-magnetization
Kawasaki runs are performed. The same material-level mean modified-GK
coupling is used in the Glauber and Kawasaki Hamiltonians, providing the
material-specific energy scale for temperatures reported in kelvin.

Constant-magnetization simulations are then evolved using Kawasaki
dynamics:\cite{Kawasaki1966} a spin $i$ is chosen at random, a spin $j$ with
$S_j = -S_i$ is chosen at random, and the pair exchange is accepted
with the Metropolis probability\cite{Metropolis1953}
\begin{equation}
    P = \min\!\left(1,\; \exp\!\left[-\frac{\Delta H}{k_B T}\right]\right).
    \label{eq:metro}
\end{equation}
The energy change for the swap is
\begin{equation}
    \Delta H = -2J\!\left[
    \sum_{\langle ik\rangle} S_i S_k
    + \sum_{\langle jl\rangle} S_j S_l\right].
    \label{eq:deltaH}
\end{equation}
The system is equilibrated for $2\times10^5$ accepted moves per
lattice site and properties are averaged over $10^6$ accepted moves
per lattice site.
We collect 1{,}000 independent configurations per temperature for ML
analysis.
Simulations are performed for eight magnetization sectors,
\[
\begin{array}{l}
M \in \{0.0,\,0.25,\,0.375,\,0.5,\\
\qquad 0.625,\,0.75,\,0.8125,\,0.875\},
\end{array}
\]
across 61 temperatures spanning the candidate-specific window set by
the Glauber $T_C$ estimate.
All candidates are simulated on three-dimensional lattices, since
each corresponds to a real crystal structure.
The EuO benchmark is treated in the same 3D workflow and serves as an
external ferromagnetic reference for graph construction,
conserved-magnetization MC, and autoencoder analysis.
The reference study uses cubic linear dimensions
$L \in \{16,20,24,32\}$ and corresponding system sizes of 4{,}096 to
32{,}768 spins.\cite{Jang2025}
For the CIF-derived lattices, $n$ denotes the integer number of
crystallographic supercell repeats, and $N$ denotes the number of
magnetic sites represented as Ising spins.
EuO and CeGaO$_3$ each contain four magnetic sites per conventional
cell, so both give $N=4n^3$. For $n=2$--$6$, this corresponds to
$N=32,\,108,\,256,\,500,$ and $864$.
CeF$_3$ contains six Ce sites per conventional cell, giving $N=6n^3$;
its Glauber range $n=5$--$9$ corresponds to
$N=750,\,1{,}296,\,2{,}058,\,3{,}072,$ and $4{,}374$.
These smaller size ladders provide multiple adjacent-size
Binder-cumulant crossings over the 61-temperature grid.
The reported Glauber transition estimates incorporate the variation
among the available adjacent-size crossings.
The subsequent Kawasaki calculations use larger material-specific
sizes: EuO $n=10,13,16,20$, corresponding to
$N=4{,}000,\,8{,}788,\,16{,}384,$ and $32{,}000$;
CeF$_3$ $n=9,11,13$, corresponding to
$N=4{,}374,\,7{,}986,$ and $13{,}182$; and
CeGaO$_3$ $n=9,10,13$, corresponding to
$N=2{,}916,\,4{,}000,$ and $8{,}788$.
The Glauber size ladders are used to establish the temperature scale
through adjacent-size Binder crossings, whereas the larger Kawasaki
sizes are used for finite-size analysis of the conserved-magnetization
phase boundaries.
For every saved configuration, the total magnetization is checked to
ensure exact conservation within the requested magnetization sector
before local affinity features are extracted for the autoencoder
analysis.

\begin{figure*}[tp]
  \centering
  \includegraphics[width=\textwidth]{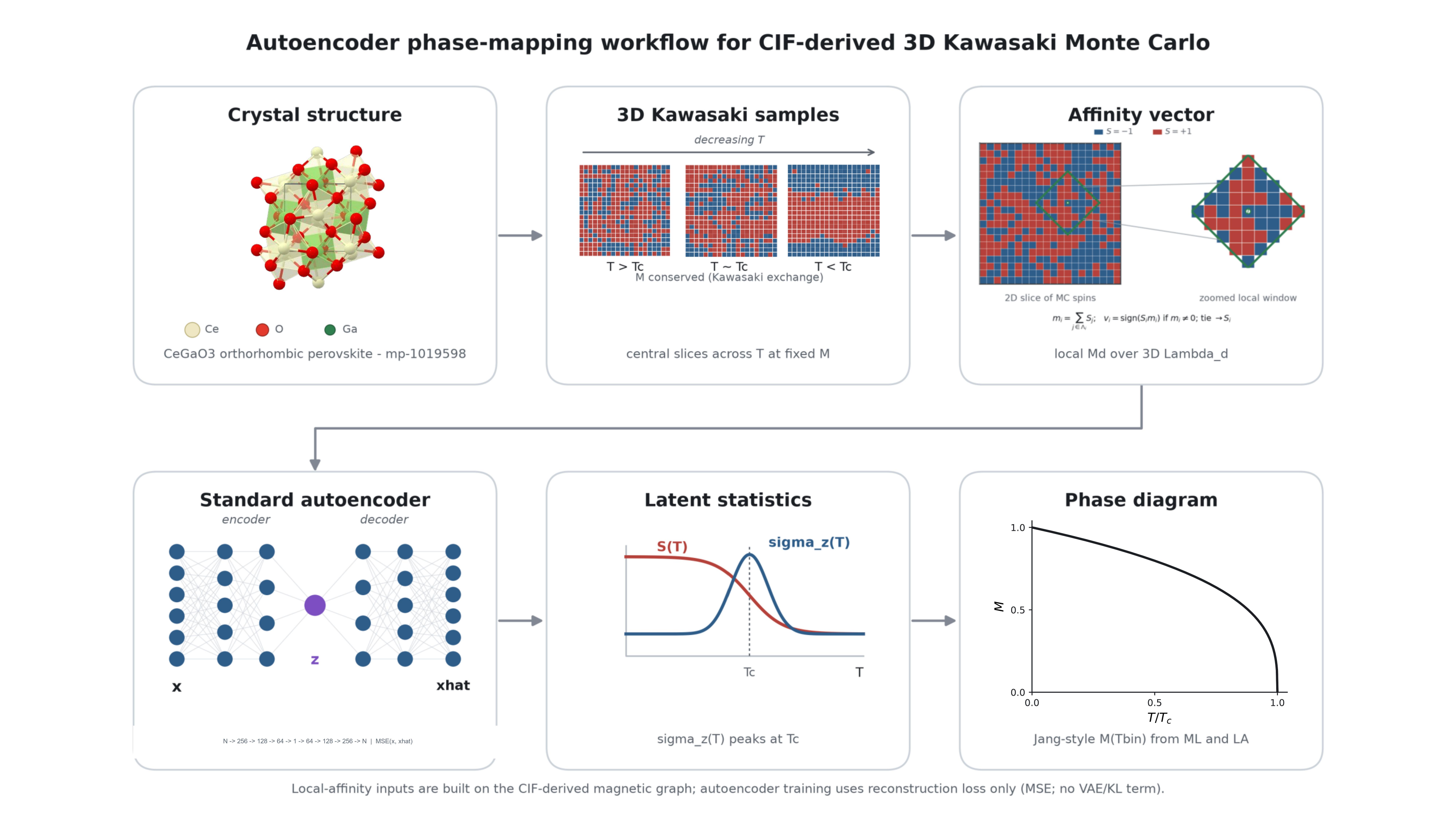}
  \caption{Kawasaki Monte Carlo and autoencoder phase-mapping workflow
    for CIF-derived three-dimensional simulations. Crystal structures
    are converted to magnetic lattices, equilibrated
    constant-magnetization Kawasaki configurations are sampled over
    temperature, local-affinity vectors are constructed from spin
    neighborhoods, and a standard autoencoder compresses these features
    to a one-dimensional latent coordinate for phase-boundary
    reconstruction.}
  \label{fig:phase_mapping_workflow}
\end{figure*}

\subsection{Local Affinity Feature and Autoencoder Analysis}

Raw spin configurations carry no explicit information about local
correlations. Following Jang and Yethiraj,\cite{Jang2025} and related
feature-construction studies,\cite{Ponte2017,Jang2022} we compute a local
affinity feature at each site $i$.  In the original square/cubic
Ising lattices, the local region $\Lambda_d$ is a diamond in 2D or a
bipyramid in 3D.  For CIF-derived magnetic graphs, the geometry-only
adaptation is to define $\Lambda_d(i)$ as the graph-distance ball on
the retained exchange network,
\[
    \Lambda_d(i)=\{j:\operatorname{dist}_G(i,j)\le d\},
\]
which contains the central site and its nearest-neighbor shell for $d=1$
and expands analogously to the window used by Jang and Yethiraj for
larger $d$. Its internal
magnetization
\begin{equation}
    M_d(i) = \sum_{j \in \Lambda_d(i)} S_j
    \label{eq:Md}
\end{equation}
is computed, and the affinity is assigned as
\begin{equation}
    v_i = \begin{cases} +1 & \text{if } M_d(i)\, S_i > 0, \\
                         S_i & \text{if } M_d(i) = 0, \\
                        -1 & \text{if } M_d(i)\, S_i < 0.
          \end{cases}
    \label{eq:affinity}
\end{equation}
The $M_d(i)=0$ case follows the tie convention of Jang and
Yethiraj\cite{Jang2025}: when the local region has zero net
magnetization, the affinity value is assigned the central spin value
rather than a separate zero class.
The graph-window size $d$ is chosen per material, system size, and
magnetization sector by maximizing the standard deviation of
$\langle v \rangle$ over the full temperature range, the same
selection criterion used by Jang and Yethiraj.

The feature matrix $\mathbf{V}$ (rows = configurations, columns =
lattice sites) is fed to the symmetric autoencoder in the final
phase-mapping workflow stage (Figure~\ref{fig:phase_mapping_workflow}).
The encoder and decoder each contain three fully connected hidden
layers with node counts $N_u^1 = 256$, $N_u^2 = 128$,
$N_u^3 = 64$ and ReLU activations; the input/output layers use a
linear activation function, consistent with the physical requirement
that the input and output representations are identical.
The latent space is one-dimensional, and its mean $S$ and standard
deviation $\sigma_S$ serve as the order parameter and its
susceptibility proxy, respectively.
Training uses the ADAM optimizer with mean squared error (MSE) loss,
a batch size of 10, a dropout rate of 0.25, and Gaussian weight
initialization with $\sigma = 0.05$.
We train for 100 epochs with a fixed ADAM learning rate
$\eta = 10^{-5}$, following the reference protocol.
Hyperparameters are summarized in Table~\ref{tab:hyperparams}.

\begin{table}[h]
  \caption{Summary of Autoencoder Hyperparameters}
  \label{tab:hyperparams}
  \begin{tabular}{lc}
    \toprule
    Hyperparameter & Value \\
    \midrule
    $N_u^1,\, N_u^2,\, N_u^3$ & 256, 128, 64 \\
    Epochs           & 100   \\
    Learning rate $\eta$ & $10^{-5}$ \\
    Dropout rate     & 0.25  \\
    Batch size       & 10    \\
    Loss function    & MSE   \\
    Optimizer        & ADAM  \\
    \bottomrule
  \end{tabular}
\end{table}

The binodal temperature $T_{\mathrm{bin}}$ is identified as the
temperature at which $\sigma_S$ is maximum.
Following the reference protocol, with $n$ in place of the reference
lattice dimension $L$, the binodal finite-size extrapolation follows
$T_{\mathrm{bin}} \sim n^{-1/\nu}$, with $\nu = 0.630$ (3D Ising,
$M=0$ critical region) and $\nu = 0.876$ (3D percolation, off-critical
sectors), determined by the Ginzburg criterion.\cite{AlsNielsen1977}
Binder-cumulant crossings and pseudocritical-temperature shifts provide
complementary finite-size checks on $\nu$.

\subsection{Effect of Activation Functions}

The choice of activation function in the autoencoder input/output
layers has a measurable effect on the accuracy of the critical
temperature estimate.
A linear activation (our default) treats the input and output layers
symmetrically, consistent with the physical requirement that the
autoencoder reconstruct its own input.
Replacing the linear with a hyperbolic tangent (tanh) activation
--- as used in some prior studies\cite{Alexandrou2020} --- introduces
an asymmetry between the input and output layers that biases the
recovered critical temperature, because the reconstruction is no
longer faithful.
This finding underscores that ML must not be applied as a black box:
domain knowledge about the symmetry of the reconstruction task is
essential for obtaining quantitatively reliable results.

% =============================================================
\section{Results and Discussion}
% =============================================================

\subsection{Candidate Selection}

The overall pipeline in Figure~\ref{fig:workflow} summarizes the
successive reduction of the candidate space and its connection to the
downstream simulation workflow. Starting from 1{,}011 Ce-containing compounds,
the out-of-fold RF stage predicts 838 ferromagnetic candidates. The
no-transition-metal-overlap criterion and CIF availability then reduce
the set to 311 structures for explicit modified-GK analysis. The
$f \!\to\! 5d$ correction retains 237 positive-exchange candidates.

Full Kawasaki MC phase-diagram calculations are computationally
expensive, so we did not simulate all 237; instead we selected two
representative Ce compounds according to two practical criteria
that make a compound both physically meaningful and amenable to
clean finite-size scaling.
First, the magnetic ordering must be driven by Ce--anion--Ce
superexchange alone, with no competing magnetic species, consistent
with the goal of discovering a genuinely cerium-based ferromagnet.
Second, the Ce coordination number must be high enough ($z \gtrsim 6$)
to produce a sharp, well-defined Binder-cumulant crossing; chain-like
or low-coordination structures were excluded on this basis.
Applying these criteria yields CeF$_3$ (mp-22070; $z = 10$--$11$,
modified-GK screening $J_{\mathrm{Ce}} = +3.66$\,meV) and
CeGaO$_3$ (mp-1019598; uniform $z = 8$,
$J_{\mathrm{Ce}} = +4.73$\,meV).
CeF$_3$ serves as the comparatively well-characterized member, while
CeGaO$_3$ represents an understudied geometry.
EuO (mp-21394) is carried through the same MC and autoencoder
analysis as an external benchmark because it is a well-established
rare-earth ferromagnet with a reported Curie temperature and a
comparatively simple magnetic sublattice.\cite{Kasuya1968,Mauger1986}
We note that the ferromagnetic labels used to train and evaluate the
RF screen derive from Materials Project spin-polarized DFT rather than
experimental measurement, so the two Ce candidates are best described
as computationally predicted ferromagnets whose Ising-limit phase
behavior this pipeline quantifies. Their suitability as uniaxially anisotropic ferromagnets
requires a subsequent magnetocrystalline-anisotropy or experimental
easy-axis validation step.

\begin{figure*}[tp]
  \centering
  \includegraphics[width=0.634\textwidth]{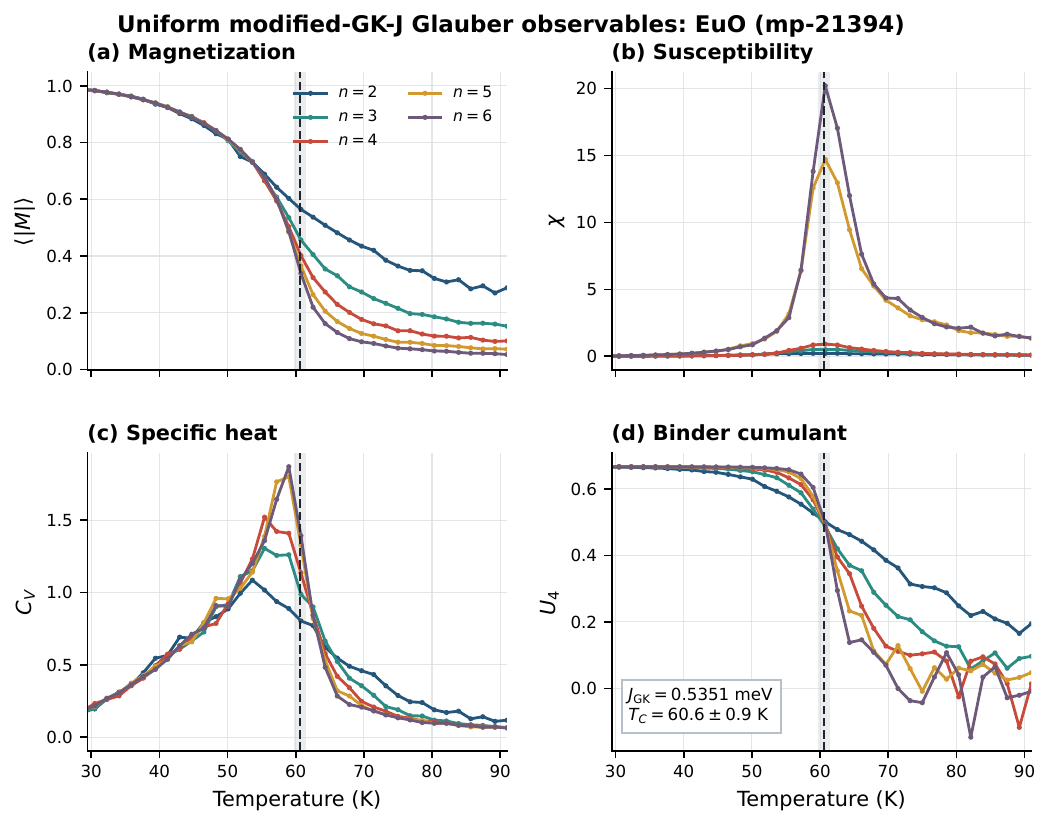}\\[0.5em]
  \begin{minipage}[t]{0.49\textwidth}
    \centering
    \includegraphics[width=0.880\linewidth]{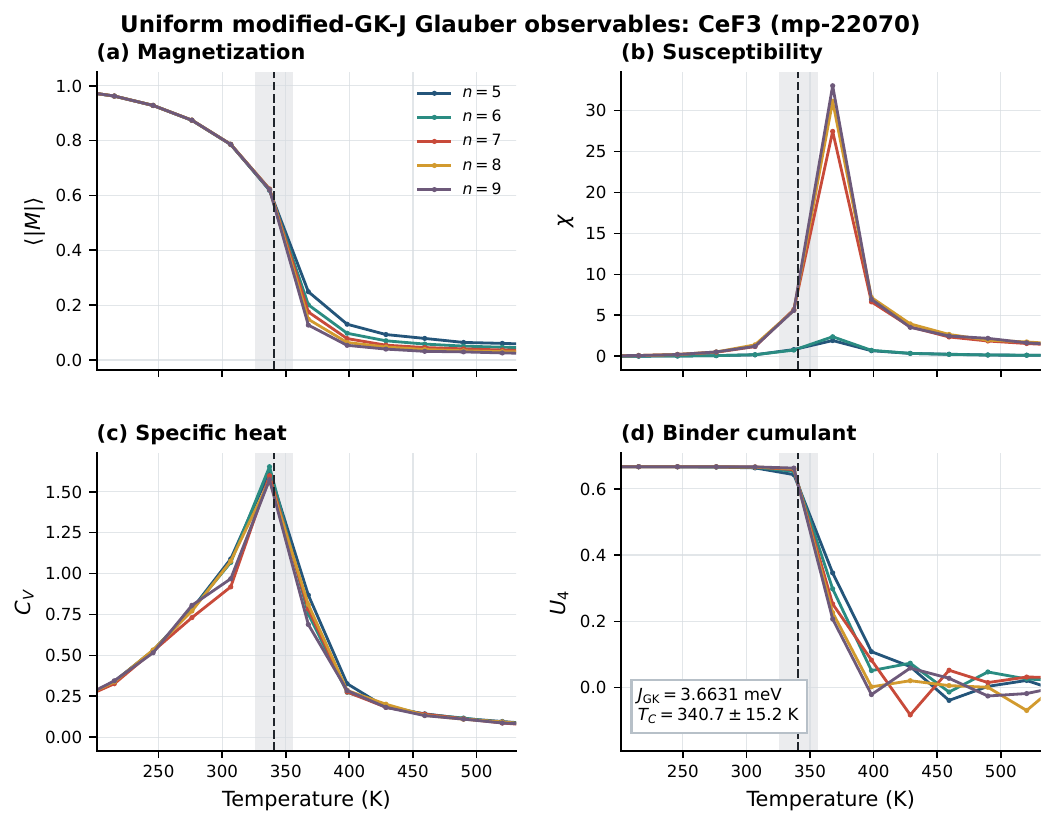}
  \end{minipage}\hfill
  \begin{minipage}[t]{0.49\textwidth}
    \centering
    \includegraphics[width=0.880\linewidth]{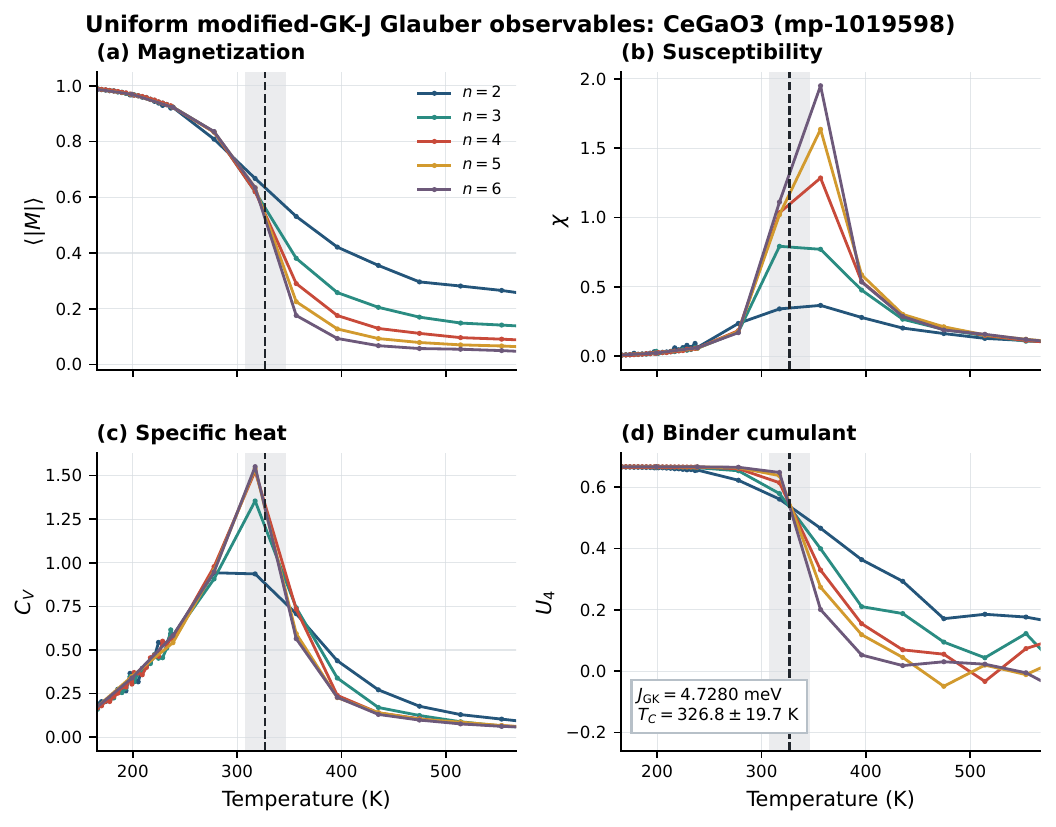}
  \end{minipage}
  \caption{Glauber thermodynamic observables obtained using the
    material-level modified-GK coupling for
    EuO (top), CeF$_3$ (bottom left), and CeGaO$_3$ (bottom right).
    Each panel set shows absolute magnetization, susceptibility,
    specific heat, and the fourth-order Binder cumulant across the
    available system sizes. Dashed lines and shaded intervals mark the
    refined adjacent-size Binder estimates: $T_C=60.6\pm0.9$\,K for
    EuO, $340.7\pm15.2$\,K for CeF$_3$, and
    $326.8\pm19.7$\,K for CeGaO$_3$. The EuO estimate lies 8.4\,K
    (approximately 12.2\%) below its experimental $T_C\approx69$\,K;
    the Ce values characterize the imposed positive uniform-$J$
    crystal-graph Ising models and are not experimentally validated
    Curie temperatures.}
  \label{fig:euo_glauber}
\end{figure*}

\begin{figure*}[tp]
  \centering
  \includegraphics[width=0.792\textwidth]{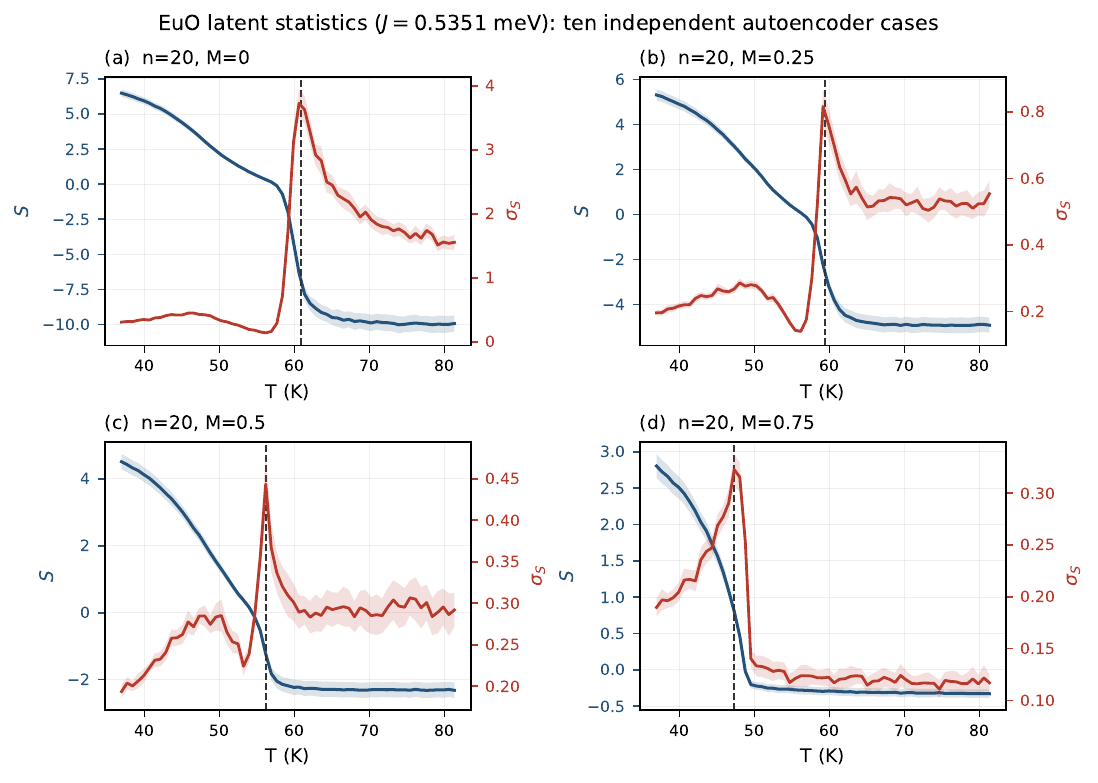}
  \caption{EuO autoencoder latent statistics for the largest simulated
    size, $n=20$, at representative conserved magnetizations. The blue
    curves show the mean latent order parameter $S$, the red curves show
    its within-temperature standard deviation $\sigma_S$, and shaded
    bands show the variation across ten independent 100-snapshot cases.
    Dashed vertical lines mark the maxima of $\sigma_S$ used to identify
    $T_{\mathrm{bin}}$. Temperatures use the modified-GK value
    $J_{\mathrm{GK}}=0.5351$\,meV.}
  \label{fig:euo_latent}
\end{figure*}

\subsection{EuO Benchmark Validation}

EuO is used here as a well-established rare-earth ferromagnetic
benchmark with a reported Curie temperature
$T_C \approx 69$\,K.\cite{Kasuya1968,Mauger1986}
It provides a reference structure for testing the same graph
construction, conserved-magnetization MC, and autoencoder analysis
applied to the Ce candidates.
The Glauber Binder-cumulant analysis using the uniform modified-GK
coupling $J_{\mathrm{GK}}=0.5351$\,meV gives
$T_C = 60.6 \pm 0.9$\,K. This is 8.4\,K
(approximately 12.2\%) below the experimental value of about 69\,K.
Figure~\ref{fig:euo_glauber} compares the associated Glauber
observables for EuO, CeF$_3$, and CeGaO$_3$; the refined transition
estimates are collected in Table~\ref{tab:glauber_gk_summary}.

\begin{table}[t]
  \caption{Glauber transition estimates using material-level
    modified-GK couplings, obtained from
    adjacent-size Binder-cumulant crossings.}
  \label{tab:glauber_gk_summary}
  \centering
  \small
  \setlength{\tabcolsep}{3.5pt}
  \begin{tabular}{@{}lccc@{}}
    \toprule
    Material & Sizes & $J_{\mathrm{GK}}$ (meV) & $T_C$ (K) \\
    \midrule
    EuO       & $n=2$--$6$ & 0.5351 & $60.6\pm0.9$ \\
    CeF$_3$   & $n=5$--$9$ & 3.6631 & $340.7\pm15.2$ \\
    CeGaO$_3$ & $n=2$--$6$ & 4.7280 & $326.8\pm19.7$ \\
    \bottomrule
  \end{tabular}
\end{table}

Finite-size extrapolation of the completed Kawasaki data using
$J_{\mathrm{GK}}=0.5351$\,meV gives the $M=0$ binodal estimates
$T_C=60.20\pm0.64$\,K from the autoencoder (AE) and
$T_C=60.59\pm0.41$\,K from the local-affinity (LA) variance.
These values characterize the present crystal-graph Ising model and
serve as an internal workflow comparison with the Glauber estimate;
they are not a replacement for the experimental EuO Curie temperature.
The same completed data set is used to quantify the phase-boundary
errors of the AE and LA methods, following the validation strategy of
Jang and Yethiraj.\cite{Jang2025}

\begin{figure*}[tp]
  \centering
  \includegraphics[width=0.880\textwidth]{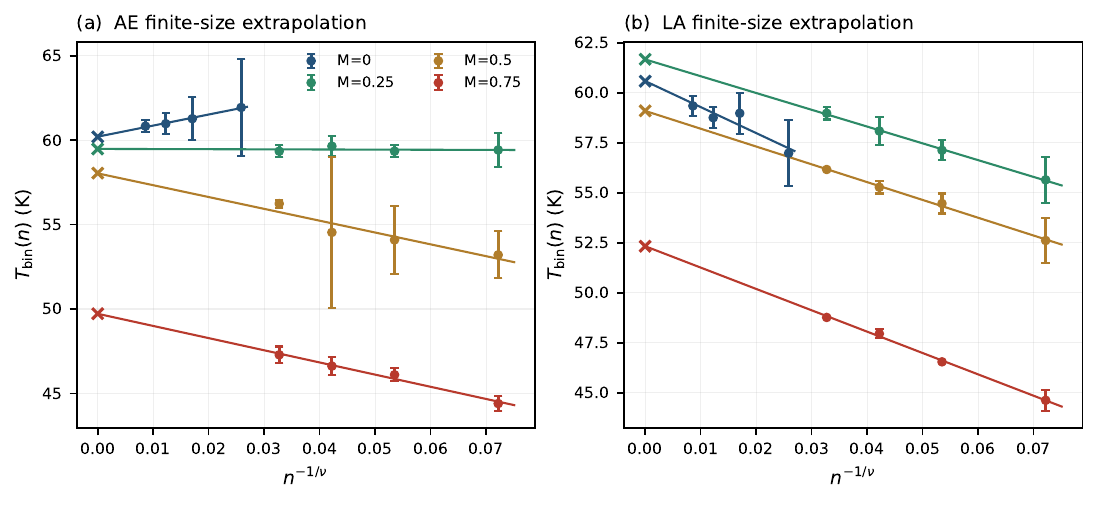}
  \caption{Finite-size extrapolation of representative EuO binodal
    sectors using (a) the autoencoder latent fluctuation and (b) the
    local-affinity variance. Points show the ten-case means at
    $n=10,13,16,$ and $20$, lines are linear fits in $n^{-1/\nu}$, and
    crosses mark the thermodynamic-limit intercepts. Following the 3D
    protocol of Jang and Yethiraj, $\nu=0.630$ is used at $M=0$ and
    $\nu=0.876$ for the off-critical sectors.\cite{Jang2025}
    Temperatures use $J_{\mathrm{GK}}=0.5351$\,meV.}
  \label{fig:FSS}
\end{figure*}

The autoencoder order parameter $S$ and its standard deviation
$\sigma_S$ are tracked as functions of temperature for each fixed
magnetization sector. Figure~\ref{fig:euo_latent} shows representative
$n=20$ results from the ten independent autoencoder cases.
The maximum in $\sigma_S$ defines $T_{\mathrm{bin}}$, and the
finite-size extrapolations in Figure~\ref{fig:FSS} give the
thermodynamic-limit binodal.

\subsection{Critical Exponent Recovery}

\begin{figure*}[tp]
  \centering
  \includegraphics[width=0.880\textwidth]{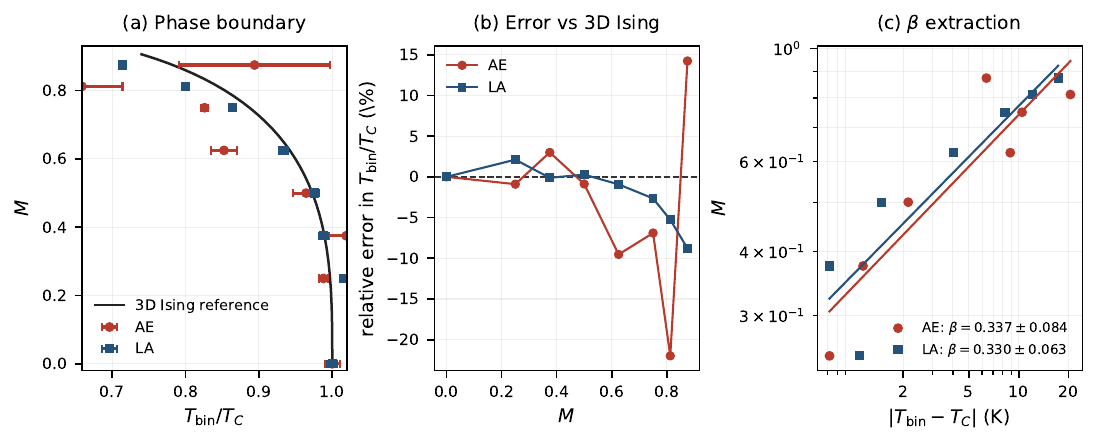}
  \caption{EuO phase-boundary reconstruction and critical-exponent
    analysis. (a) Thermodynamic-limit AE and LA binodals compared with
    the 3D Ising reference curve.\cite{Talapov1996} (b) Relative errors
    of the reconstructed reduced binodal temperatures. (c) Logarithmic
    fits of $M$ versus $|T_{\mathrm{bin}}-T_C|$ used to extract the
    order-parameter exponent $\beta$.}
  \label{fig:phasediagram}
  \label{fig:exponents}
\end{figure*}

After $T_{\mathrm{bin}}$ is extracted across magnetization sectors,
the order-parameter exponent $\beta$ is obtained from a
logarithmic plot of $M$ versus $|T_{\mathrm{bin}} - T_C|$,
where $M \sim |T_{\mathrm{bin}} - T_C|^\beta$.
For EuO, the completed analysis gives
$\beta_{\mathrm{AE}}=0.337\pm0.084$ and
$\beta_{\mathrm{LA}}=0.330\pm0.063$, both consistent within uncertainty
with the 3D Ising value $\beta\approx0.326$.
The mean absolute phase-boundary errors relative to the 3D Ising
reference are $7.18\%$ for AE and $2.51\%$ for LA. The larger AE
deviations at high magnetization are retained in
Figure~\ref{fig:phasediagram} rather than filtered from the analysis.

The present analysis reports the order-parameter exponent $\beta$ from
the phase boundary. The finite-size thermodynamic observables generated
by the Glauber stage provide routes to estimating the remaining critical
exponents:
the specific-heat exponent $\alpha$ from
$C \sim |T-T_C|^{-\alpha}$ or from the finite-size scaling of
$C_{\max}(L)$, the susceptibility exponent $\gamma$ from
$\chi \sim |T-T_C|^{-\gamma}$ or $\chi_{\max}(L)$, and the correlation
length exponent $\nu$ from Binder-cumulant crossings and the
finite-size shift of pseudocritical temperatures.
Equivalently, the finite-size scaling relations
$M(T_C,L)\sim L^{-\beta/\nu}$,
$\chi_{\max}(L)\sim L^{\gamma/\nu}$, and
$C_{\max}(L)\sim L^{\alpha/\nu}$ provide cross-checks on the exponent
set.
Because all simulations are performed with a three-dimensional
coarse-grained Ising Hamiltonian on crystal-derived lattices, the
3D Ising exponent set is the appropriate reference for this modeling
framework.
Agreement between the extracted exponents and the 3D Ising reference
values provides a final check that the thermodynamic and
autoencoder-derived observables encode the physical critical behavior
rather than spurious correlations. The completed EuO $\beta$ extraction
is summarized in Figure~\ref{fig:exponents}.

\subsection{Ce-Based Candidate Phase Diagrams}

\begin{figure*}[b]
  \centering
  \includegraphics[height=0.300\textheight,keepaspectratio]{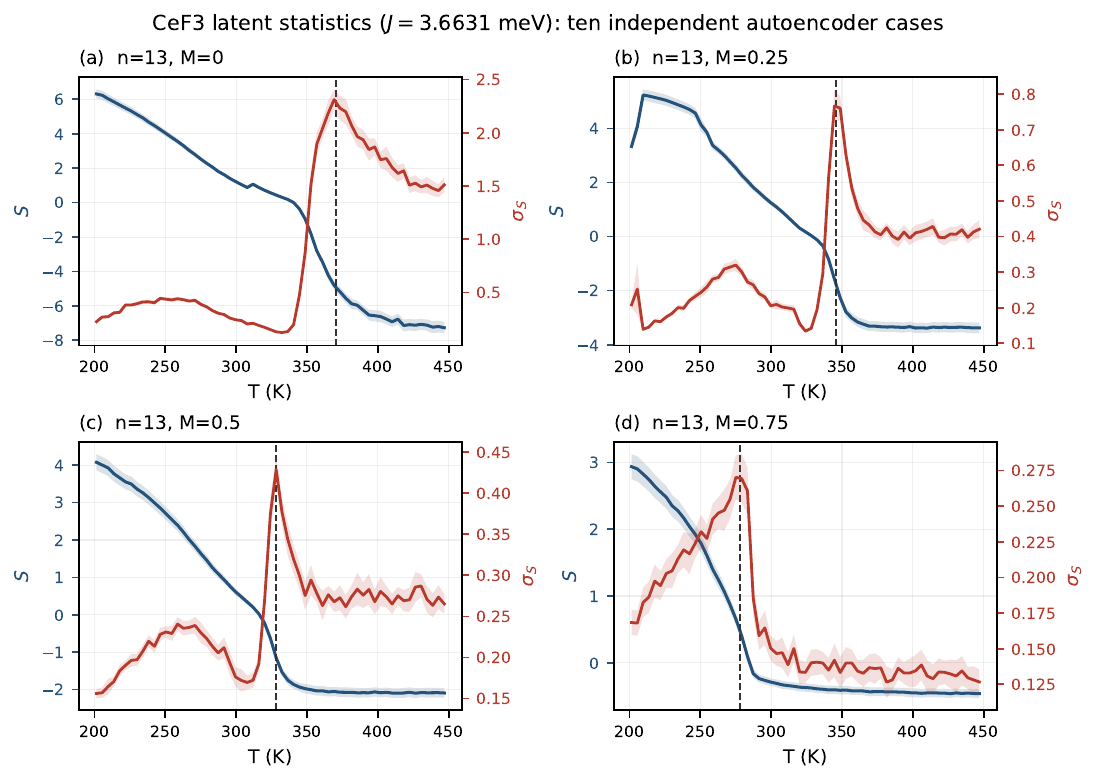}\\[-0.4em]
  \includegraphics[height=0.300\textheight,keepaspectratio]{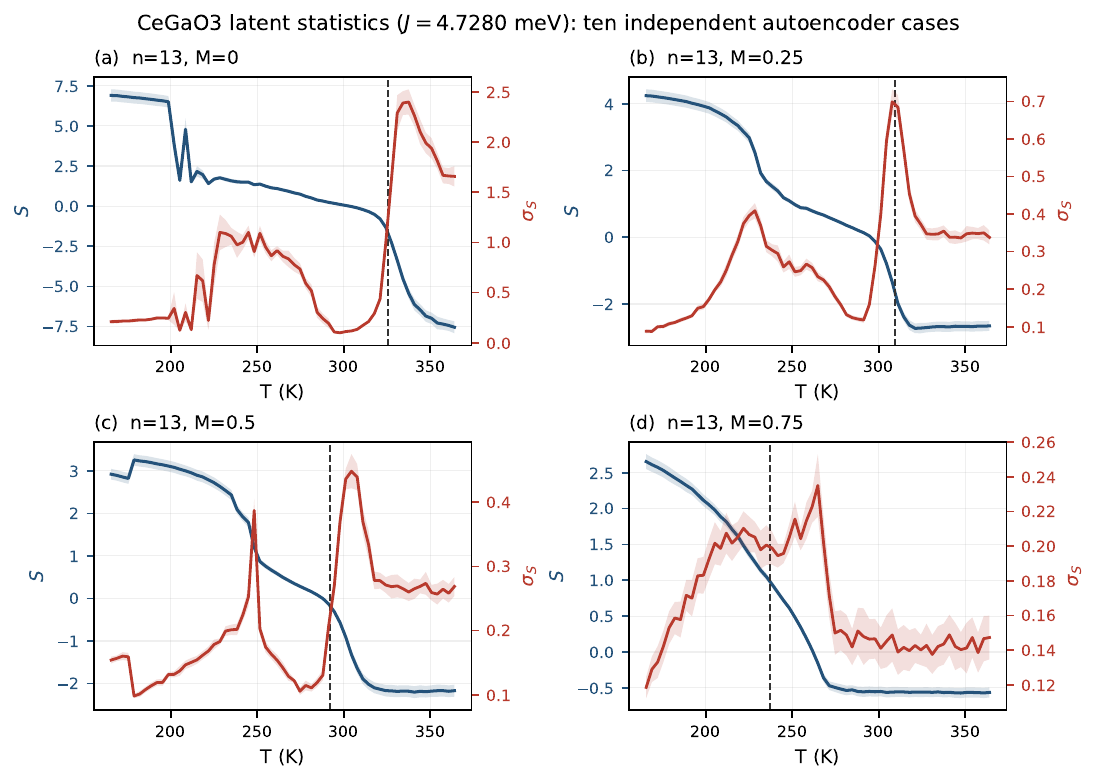}
  \caption{Latent statistics from the Jang-style autoencoder analysis
  for CeF$_3$ (top) and CeGaO$_3$ (bottom). The latent order parameter
  $S$ and its standard deviation $\sigma_S$ are shown for representative
  fixed-magnetization sectors; the peak in $\sigma_S$ defines the
  corresponding finite-size transition estimate.}
  \label{fig:ce_latent}
\end{figure*}

The three-size analysis uses CeF$_3$ lattice sizes $n=9,11,13$ and
CeGaO$_3$ lattice sizes $n=9,10,13$. Figure~\ref{fig:ce_latent} shows
the latent order parameter and latent standard deviation for
representative fixed-magnetization sectors. Both materials exhibit
temperature-dependent maxima in $\sigma_S$, from which the finite-size
transition temperatures are obtained.

\begin{figure*}[tp]
  \centering
  \includegraphics[width=0.880\textwidth]{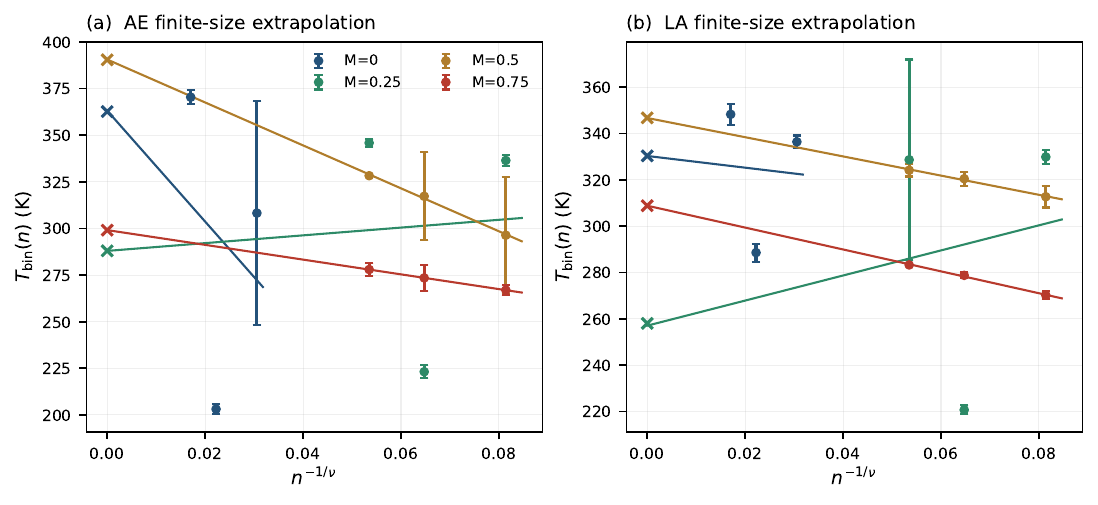}\\[-0.3em]
  \includegraphics[width=0.880\textwidth]{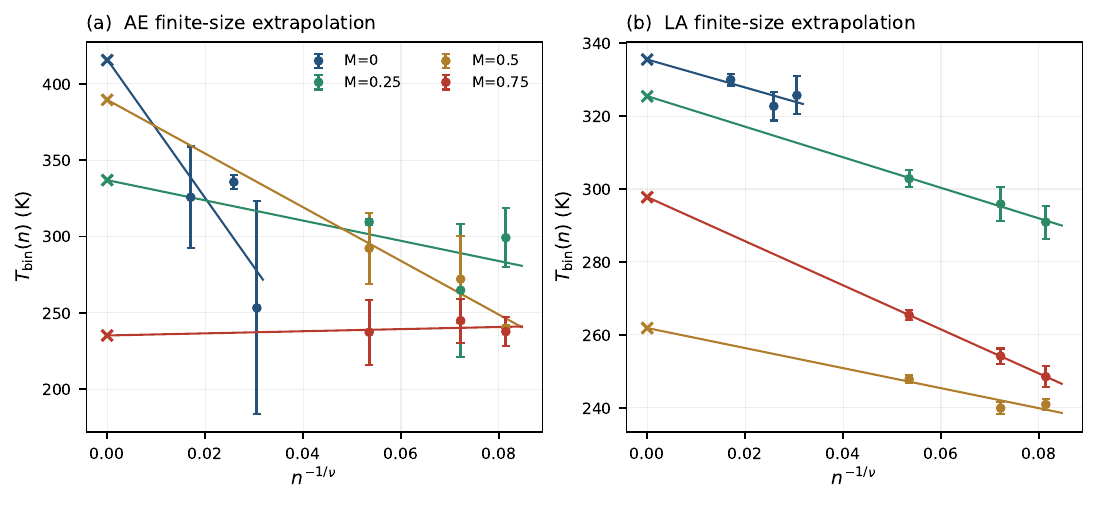}
  \caption{Three-size finite-size-scaling analysis for CeF$_3$ (top;
  $n=9,11,13$) and CeGaO$_3$ (bottom; $n=9,10,13$). Autoencoder and
  local-affinity transition estimates are extrapolated independently
  for each magnetization sector. Including larger lattice sizes could
  provide additional constraints on the scaling fits and reduce their
  sensitivity to any single size.}
  \label{fig:ce_fss}
\end{figure*}

The finite-size transition estimates are shown in
Figure~\ref{fig:ce_fss}. For CeGaO$_3$, the reconstructed phase
boundary gives $\beta_{\mathrm{LA}}=0.363\pm0.057$ and
$\beta_{\mathrm{AE}}=0.300\pm0.209$. The corresponding mean absolute
phase-boundary errors are $5.37\%$ for the local-affinity analysis and
$19.68\%$ for the autoencoder analysis. Ferromagnetic order in
CeGaO$_3$ has also been predicted by first-principles
calculations,\cite{Kabi2025} and both exponents recovered here lie
close to the three-dimensional Ising value $\beta\approx
0.326$,\cite{Campostrini1999} with the local-affinity estimate agreeing
within uncertainty. Taken together, these results identify CeGaO$_3$ as
a candidate Ising ferromagnet and motivate further study of this
compound.

\begin{figure*}[b]
  \centering
  \includegraphics[width=0.880\textwidth]{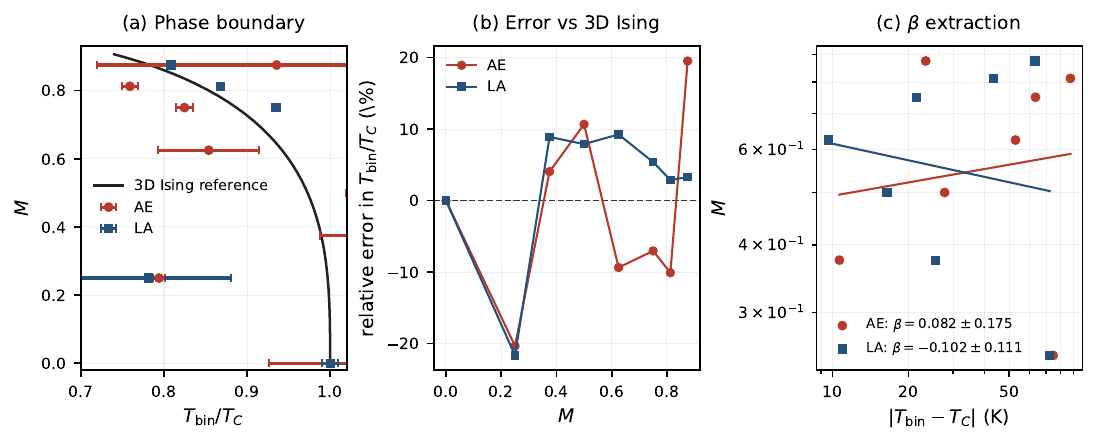}\\[-0.3em]
  \includegraphics[width=0.880\textwidth]{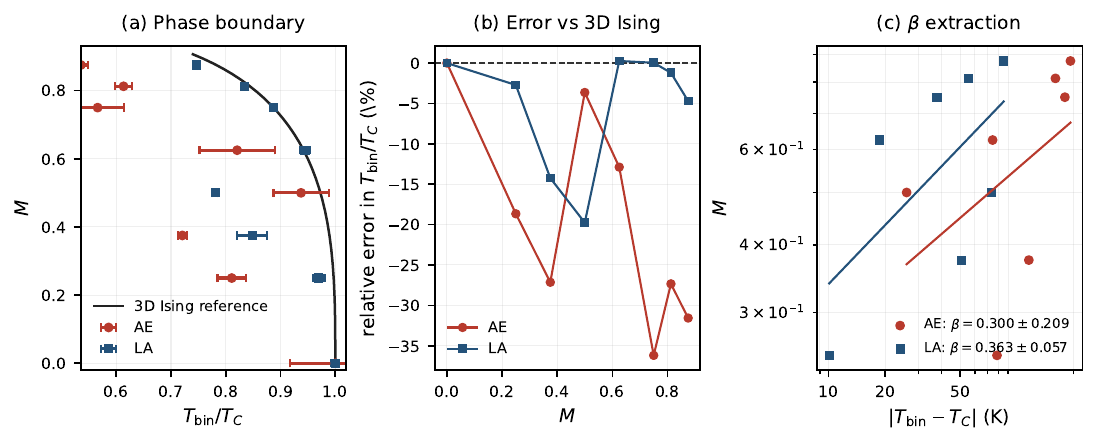}
  \caption{Reconstructed phase boundaries and critical-region fits for
  CeF$_3$ (top) and CeGaO$_3$ (bottom), compared with the
  three-dimensional Ising reference. Autoencoder and local-affinity
  results are shown for each material.}
  \label{fig:ce_phase_beta}
\end{figure*}

For CeF$_3$, the fits give $\beta_{\mathrm{AE}}=0.082\pm0.175$ and
$\beta_{\mathrm{LA}}=-0.102\pm0.111$. The mean absolute
phase-boundary errors are $10.14\%$ for the autoencoder analysis and
$7.41\%$ for the local-affinity analysis. Although our model predicts
CeF$_3$ to be ferromagnetic, its magnetic behavior remains debated: the
strong exchange obtained here reflects the $4f$--$5d$ virtual-excitation
channel, yet the recovered exponents are far from the Ising value and,
in the local-affinity case, are not physically meaningful for an
order-parameter exponent. This is consistent with a picture dominated
by magnetic excitations rather than by a sharp Ising transition, and no
distinct magnetic ordering was reported in the most recent single-crystal
study of CeF$_3$.\cite{Savinkov2016} The reconstructed phase boundaries
and exponent fits for both materials are shown in
Figure~\ref{fig:ce_phase_beta}.

\clearpage
% =============================================================
\section{Concluding Remarks}
% =============================================================

We have presented a multi-stage machine learning pipeline for the
systematic discovery and computational phase-diagram characterization
of Ce-based ferromagnetic candidates, with the longer-term materials
target being compounds that also exhibit strong uniaxial magnetic
anisotropy.
The key innovations are: (1) a Random Forest structural-screening stage
adapted from the structural-feature-plus-SOAP descriptor strategy of
Broyles and colleagues\cite{Broyles2024} for defining the Ce
ferromagnetic candidate pool; (2) a Ce-specific $f \!\to\! 5d$
correction to the GK superexchange screening step for selecting
positive-exchange Ce candidates; (3) Glauber MC for critical-temperature estimation that sets
the absolute temperature window in kelvin; (4) constant-magnetization
Kawasaki MC phase-diagram calculations on material-derived magnetic
graphs; and (5) an unsupervised
    autoencoder that maps spin-configuration data to a one-dimensional
    latent order parameter and reconstructs the phase diagram with
    quantified binodal errors.

The EuO benchmark provides an external ferromagnetic reference for
testing the simulation and autoencoder workflow.
The two Ce candidates --- CeF$_3$ and CeGaO$_3$ --- remain
cerium-based positive-exchange ferromagnetic candidates. Their
three-size autoencoder and local-affinity analyses produce the
phase-boundary and critical-exponent results reported here.
They should not yet be interpreted as experimentally established
uniaxial ferromagnets; that designation requires direct confirmation
of easy-axis magnetic anisotropy.

Several open directions remain.
First, the present simulations employ a single effective exchange
parameter $J_{\mathrm{eff}}$ obtained from the average exchange
interaction predicted by the modified Goodenough--Kanamori screening
model. This effective coupling is applied uniformly to all retained
magnetic interactions in the Ising Hamiltonian.
Assigning distinct pathway-resolved exchange parameters from the GK
model or from DFT+$U$ remains a future extension of the Hamiltonian.
Second, DFT-based confirmation of the predicted ferromagnetic
ground states, including magnetocrystalline-anisotropy energies and
easy-axis directions, would strengthen the case for experimental
synthesis, since the underlying Materials Project labels are themselves
DFT-derived rather than experimentally measured.
The framework is general and can be extended beyond Ce compounds to
broader rare-earth and actinide magnetic systems, where analogous
$f$-orbital corrections to the GK rules are expected to play a similar
role, providing a scalable approach for AI-assisted magnetic materials
discovery.

% =============================================================
%  ASSOCIATED CONTENT
% =============================================================

\begin{acknowledgement}
The authors thank the Oklahoma State University High Performance
Computing Center (Pete cluster) and the OU Supercomputing Center
for Education and Research (OSCER/Schooner cluster) for computational
resources.
This work was supported by the University of Central Oklahoma, its
School of Engineering, and its Office of High-Impact Practices (OHIP)
through an RCSA grant.
\end{acknowledgement}

\section*{Supporting Information, Data, and Code Availability}
Supporting analysis materials, data-processing scripts, the modified-GK
screening workflow, Monte Carlo simulation code, affinity-feature
extraction, and autoencoder analysis code will be made publicly available
on GitHub. The accompanying materials include details of the SOAP descriptor
parameterization, Random Forest hyperparameter tuning, the Ginzburg
criterion analysis for determination of the critical region, binodal
comparisons using different finite-size scaling exponents, and
principal component analysis of the latent space.

% =============================================================
%  REFERENCES
% =============================================================

\end{document}